# Thin Films on the Skin, but not Frictional Agents, Attenuate the Percept of Pleasantness to Brushed Stimuli

Merat Rezaei, *Student Member, IEEE*, Saad S. Nagi, Chang Xu, *Student Member, IEEE*, Sarah McIntyre, Håkan Olausson, Gregory J. Gerling, *Senior Member, IEEE*

*Abstract–* Brushed stimuli are perceived as pleasant when stroked lightly on the skin surface of a touch receiver at certain velocities. While the relationship between brush velocity and pleasantness has been widely replicated, we do not understand how resultant skin movements – e.g., lateral stretch, stick-slip, normal indentation – drive us to form such judgments. In a series of psychophysical experiments, this work modulates skin movements by varying stimulus stiffness and employing various treatments. The stimuli include brushes of three levels of stiffness and an ungloved human finger. The skin's friction is modulated via non-hazardous chemicals and washing protocols, and the skin's thickness and lateral movement are modulated by thin sheets of adhesive film. The stimuli are hand-brushed at controlled forces and velocities. Human participants report perceived pleasantness per trial using ratio scaling. The results indicate that a brush's stiffness influenced pleasantness more than any skin treatment. Surprisingly, varying the skin's friction did not affect pleasantness. However, the application of a thin elastic film modulated pleasantness. Such barriers, though elastic and only 40 microns thick, inhibit the skin's tangential movement and disperse normal force. The finding that thin films modulate affective interactions has implications for wearable sensors and actuation devices.

## I. INTRODUCTION

We commonly give and receive touch with others in affective social and emotional interactions. For instance, a caress of another's forearm might provide comfort while in distress, a hug from a loved one might signal remorse or help reestablish a long-awaited connection, and a series of taps and pats might signal gratitude or attention. In these types of affective exchange, the receiver judges emotional valence of the communication, which might be signaled by many interrelated physical factors [1].

Within the field of affective touch, the percept of 'pleasantness' is typically studied by delivering soft brush stimuli to the skin of human volunteers, who evaluate the touch they receive [2]–[4]. In addition to brush stimuli, human touch is similarly perceived as pleasant when likewise delivered slowly at low forces, and may help suppress pain and negative emotions [5]–[8]. Typically, a soft brush is stroked along the skin of the dorsal forearm at forces about 0.2 to 0.4 N and velocities between 0.1 and 30 cm/s [3]. Psychophysical evaluation shows that, at a group level, the velocity of the stimulus modulates pleasantness in a relationship that resembles an inverted U-shaped curve, with the greatest pleasantness reported at velocities between 1 and 10 cm/s [2], [9]. Both robot controlled and human delivered brushing has produced similar results [10]. Additional efforts have considered distinct body sites, brushes with textured surfaces (e.g., velvet, burlap, cotton, denim), ties to affiliative bonds and social cognition, and inter- versus intra-personal touch, but none have inquired into modulation of the mechanical properties of contact [4], [11]–[14].

Aside from the impact of brush velocity, we do not understand the nature of the resultant skin movements that drive our judgment of pleasantness. For instance, a brush stroke stretches the skin laterally, generates a range of forces and force rates, vibrational waves upon contact, and stick-slip events. Such interactions could drive observed firing patterns in certain afferent subtypes, such as C-tactile afferents' preference for 1-10 cm/s stroking velocities, as opposed to Aβ afferents' linearly increasing firing rate with velocity. Further, high-threshold mechanoreceptors do not respond to a soft brush, but they do respond to a rough (stiffer) brush [15]. At present, we do not understand the origin of such signaling differences, which could be related, in part, to skin mechanics.

Most efforts to directly quantify the deformation and stretch of the skin have focused on contact with transparent glass or elastomer surfaces [16], [17]. Other approaches have imaged contact interactions between human touchers and receivers, though neither for brushing stimuli nor local states of stress [18]. For non-transparent, brush stimuli, visualizing skin movement is particularly difficult. Moreover, placing a sensor or barrier on the receiver's skin changes the nature of the contact interaction. Therefore, approaches using microphones have sought to analyze audible output resulting from skin contact [19]. Furthermore, various engineered devices have sought to produce social touch [20]. In experiments focusing on the pleasantness of performing active touch, as opposed to its passive receipt, various frictional agents have been applied to the skin [21] as well as emollients [22]. Such efforts seek to perturb contact interactions at the skin surface.

This work describes psychophysical experiments to modulate skin movements and evaluate their impact on pleasantness. In contrast to measurements between the brush and skin, our distinct approach 1) varies stimulus properties, by using brushes of distinct bristle stiffness and the human finger, and 2) utilizes skin treatments to isolate attributes of adhesion, friction, film thickness, and lateral mobility.

M. Rezaei, C. Xu, and G. J. Gerling are with the School of Engineering and Applied Science, at the University of Virginia, USA (e-mail: {mr3wq, gg7h}@virginia.edu).

S. S. Nagi, S. McIntyre and H. Olausson are with the Center for Social and Affective Neuroscience (CSAN), Linköping University, Sweden.

## II. METHODS

### A. Stimuli and Skin Treatments

Three brushes were employed with increasing levels of bristle stiffness, Fig. 1A, named 'smooth,' 'hybrid,' and 'rough.' The smooth brush is made of goat hair, similar to those used in prior efforts [2], [3]. The hybrid brush is made of coarser pig hair. The rough brush is made of stiff, synthetic plastic. All brushes were 5 cm wide. The fourth stimulus, only used in Experiment 2, was the ungloved finger, which was marked at a length of 5 cm to maintain about the same contact width as the brushes.

Several treatments were used to alter the properties of the skin across the psychophysical experiments, Table 1. In Experiment 1, a thin film (Tegaderm, 3M, Part 1626W, 40 microns thick, adhesive on one side, 10 by 12 cm), calamine spray (CVS Calamine Plus, active ingredient calamine 8%), and an emollient lotion (Vaseline Advanced Repair) were used. An example application of Tegaderm film on the forearm of one participant is shown in Fig. 1B. Tegaderm film, calamine spray, and emollient lotion create a direct barrier between skin and stimulus, stiffen the skin, and smoothen the skin, respectively. In Experiment 2, hyaluronic acid (Cosmedica Skincare, humectant, main ingredients: distilled water, sodium hyaluronate, benzylalcohol-DHA), room temperature water (washed skin, then patted dry), and soap (washed skin, then patted dry, main ingredient: sodium tallowate) were used. Hyaluronic acid and water increase hydration and therefore friction, and soap decreases friction, as detailed further in *Section III.B*. In Experiments 3, 4, and 5, distinct configurations of Tegaderm film were used to decouple attributes of skin adhesion, film thickness, and friction. Configurations included two layers applied on top of each other (adhesive, 80 microns thick), two layers folded over each other (non-adhesive, 80 microns thick), one layer (9 cm length by 5 cm width), and one layer (6 cm length by 5 cm width).

### B. Participants

Thirty-four participants, balanced roughly by gender, ages 18-35, were recruited across all experiments, with n=14 in Experiment 1, and n=5 in each of Experiments 2, 3, 4, and 5, respectively. No participant was used in more than one experiment to avoid potential biases. The study was approved by the local institutional review board, with informed consent obtained from all participants.

### C. Experimental Procedures

Each participant was seated on the opposite side of a curtain from the trained experimenter, who delivered stimuli by hand using published protocols [2], Fig. 1C. The same site on a participant's dorsal forearm was used for every trial, except when a skin treatment might cause lingering or skin property-changing effects. For example, hyaluronic acid changes the skin's friction. In such situations, both arms of a participant were used interchangeably. In particular, this was the case between conditions in Experiments 1 and 2 of calamine/emollient, hyaluronic acid/water, and hyaluronic acid/soap. In contrast, Tegaderm can leave a tingling sensation when detached from the skin, so participants were given a 5-minute break upon its removal, or a duration necessary for this sensation to cease. The order of the stimuli was selected per trial by a custom computer program which randomized the brush velocity and brush stiffness. The treatment order was counterbalanced between participants.

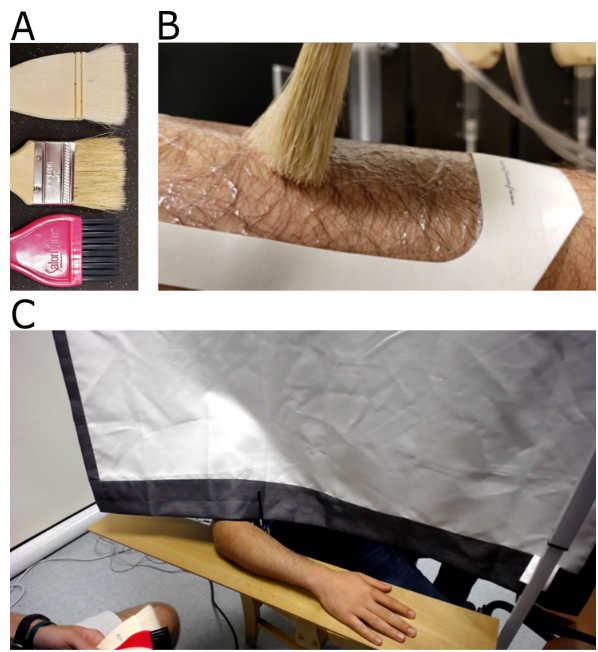

**Figure 1: Experimental setup. (A)** Brush stimuli increasing in stiffness from top to bottom were presented in randomized order, under various skin treatments, including **(B)** Tegaderm film applied to the dorsal forearm. **(C)** Participants were separated from the experimenter by a curtain and asked to rate stimulus pleasantness per trial using a visual analog scale, ranging from 'very unpleasant' to 'very pleasant.'

To reduce variability in delivering the stimuli, the angle of contact between stimulus and skin was kept at 90 degrees, while its normal force was delivered at about 0.4 N [2], [3]. The velocities delivered were a subset of 1, 3, 10, and 30 cm/s, varying by experiment. The experimenter who delivered the stimuli practiced the technique beforehand against a high resolution, pressure sensitive mat (TactArray Sensor, PPS, Hawthorne, CA, USA) to become consistent at delivering this force over the full length of the stroke.

TABLE 1: OVERVIEW OF CONDITIONS USED IN EACH PSYCHOPHYSICAL EXPERIMENT

|  | **Experiment 1** | **Experiment 2** | **Experiment 3** | **Experiment 4** | **Experiment 5** |
|---|---|---|---|---|---|
| **Stimulus** | Smooth, Hybrid, Rough Brush | Smooth, Hybrid, Human Finger | Smooth, Rough Brush | Smooth, Rough Brush | Smooth, Hybrid Brush |
| **Skin Treatment** | Tegaderm, Calamine Spray, Emollient | Hyaluronic Acid, Water, Soap | Tegaderm, Folded Tegaderm, 2xTegaderm | Tegaderm | Tegaderm, 9 cm Hole, 6 cm Hole |
| **Velocities** | 1, 3, 10, 30 cm/s | 1, 3, 30 cm/s | 10 cm/s | 10 cm/s | 1, 3, 10 cm/s |
| **Number of Participants** | 14 | 5 | 5 | 5 | 5 |

After each trial, participants were asked to rate pleasantness using a graphical user interface with a visual analog scale from 'very unpleasant' to 'very pleasant' with blind values of -5 to 5 [23].

### III. EXPERIMENTS AND RESULTS

Five psychophysical experiments were performed, as outlined in Table 1. Their procedures and results are given below. In addition, an instrumented, simulated skin was used to evaluate the force rates delivered across brush stiffness.

#### A. Experiment 1

*Procedures.* Three brush stiffness stimuli were employed under four skin conditions: 1) untreated skin; 2) direct barrier (Tegaderm film); 3) stiffened skin (calamine spray); and 4) smoothed skin (emollient lotion). See *Section II.A* for exact product numbers. Four brush velocities employed were 1, 3, 10, and 30 cm/s.

*Results.* A three-way repeated measures ANOVA was conducted to determine the effects of brush, velocity, and skin treatment on pleasantness. All three factors significantly affected pleasantness ratings, with the largest effect from brush (brush: $F = 49.27$, $p < 0.05$, $\eta^2 = 0.57$; skin treatment: $F = 13.0$, $p < 0.05$, $\eta^2 = 0.07$; velocity: $F = 3.4$, $p < 0.05$, $\eta^2 = 0.007$), Fig. 2A-C. Post-hoc contrast tests reveal that only Tegaderm had a significant effect compared to untreated skin, with an overall improvement of 0.757 ($p < .001$), whereas calamine and emollient did not. An increase in brush stiffness consistently decreased pleasantness with no overlap in 95% confidence intervals (smooth: [1.42, 2.68]; hybrid: [-0.54, 0.73]; rough: [-2.6, -1.34]).

#### B. Experiment 2

*Procedures.* To alter the frictional properties of the skin due to the hydration of the stratum corneum [24], other non-impedimentary skin treatments were introduced. Three treatments were selected, including hyaluronic acid (a humectant), and water (washed skin, then patted dry) to increase hydration and therefore friction, and soap (washed, patted dry, main ingredient: sodium tallowate) to decrease friction. We expect untreated skin, hyaluronic acid, water, and soap treatments to yield coefficients of kinetic friction of 0.45-0.65, 1.05-2.62, 0.7-1.0, and <0.45, respectively [24]. Given the rough brush had such a significant impact on pleasantness in Experiment 1, which might override any effect of a skin treatment, we focused Experiment 2 on the smooth and hybrid brushes, while introducing the human finger for comparison.

*Results.* Fig. 2D-F shows that even large changes to the surface friction of skin incite little if any change in perceived pleasantness ($F = 0.4$, $p > 0.5$, $\eta^2 = 0.003$), with an overlap of 95% confidence intervals for all skin treatments. Post-hoc contrast tests showed no significant difference in pleasantness compared to untreated skin for any of the skin treatments. This is observed across all brush stimuli. On another note, the pleasantness of the finger as the stimulus (CI [-0.02, 3.07]) was similar to that of the smooth brush (CI [-0.85, 3.93]).

#### C. Experiment 3

*Procedures.* To further analyze the various coupled attributes that Tegaderm film might induce, three factors decoupled included skin adhesion, film thickness, and frictional change. Film thickness and adhesion were varied by using one sheet of Tegaderm (40 microns thick), two stacked sheets of Tegaderm with one adhesive side ('2xTegaderm,' 80 microns thick), and two stacked sheets of Tegaderm with no adhesive side ('Folded Tegaderm', 80 microns thick), held in place with thin strips of tape on the edges. Only the smooth and rough brushes were evaluated, and at a single velocity.

*Results.* As observed for Experiment 1, pleasantness decreased for the rough brush (CI [-1.86, 0.47]), Fig. 2G-H. Likewise, for the rough brush, each of the Tegaderm configurations modulate pleasantness with an increase to a more neutral value. A two-way, repeated measures ANOVA shows no significant effect on pleasantness ratings by skin treatments, in contrast to the stimuli ($F = 16.1$, $p < 0.05$). The Tegaderm configurations do not exhibit significant differences compared to each other. However, with the smooth brush (CI [1.32, 3.65]), the 'Folded Tegaderm' (CI [-0.4, 1.4]) case with no adhesive side impeded pleasantness compared to 'Tegaderm' (CI [0.58, 2.38]) and '2xTegaderm' (CI [0.33, 2.13]) that adhere to the skin. After conducting post-hoc contrast tests, no difference was observed between the adhesive Tegaderm configurations and the 'Normal' (CI [-0.55, 1.25]) case in this experiment ($p > 0.05$), as had been observed in Experiment 1, thus leading into Experiment 4, which directly investigated the use of one sheet of Tegaderm.

#### D. Experiment 4

*Procedures.* A direct comparison was made between the 'Normal' untreated skin and 'Tegaderm' applied cases, for smooth and rough brushes. Only a single velocity was tested, at 10 cm/s. The reasoning behind running this experiment is detailed in the *Results* of Section 3.C.

*Results.* In the absence of skin treatments other than just a single layer of Tegaderm, the results remained consistent with Experiment 1 for the smooth brush (CI [1.03, 4.74]), Fig. 2I-J. The pleasantness of the rough brush (CI [-1.88, 1.83]) was only slightly more neutral than unpleasant, as in Experiment 1. This could be due to sample size limitations, or may indicate that absolute values of pleasantness are not comparable between experiments with unique skin treatments and stimulus factors.

#### E. Experiment 5

*Procedures.* The impact of modulating the skin's lateral motion on pleasantness was investigated by varying rectangular hole sizes in the Tegaderm of 6 cm and 9 cm lengths and 5 cm width, using the smooth and hybrid brushes, Fig. 2K-L. The level of lateral mobility in the skin was hypothesized to decrease in the order of 'Normal', '9 cm Hole', '6 cm Hole', and 'Tegaderm' respectively. To maintain a consistent stroke length and contact duration, all brush strokes were made at a 6 cm length. Brush strokes were executed at 1, 3, and 10 cm/s.

*Results.* As with the other experiments, the smooth brush was more pleasant than the hybrid brush, (linear contrast estimate, -1.68, $p < 0.05$). The effects of 'Tegaderm' are consistent with those of Experiment 1 in the attenuation of pleasantness across brushes (linear decrease in pleasantness: normal: -2.32, Tegaderm: -0.17). However, the use of Tegaderm film with a hole played no role, compared to the normal non-Tegaderm film condition (6 cm hole: -2.13; 9 cm hole: -2.11). This further suggests that the presence of a direct barrier at the contact interface, along with the stiffness of the stimulus, impact pleasantness more than modifications to the skin's friction or stiffness.

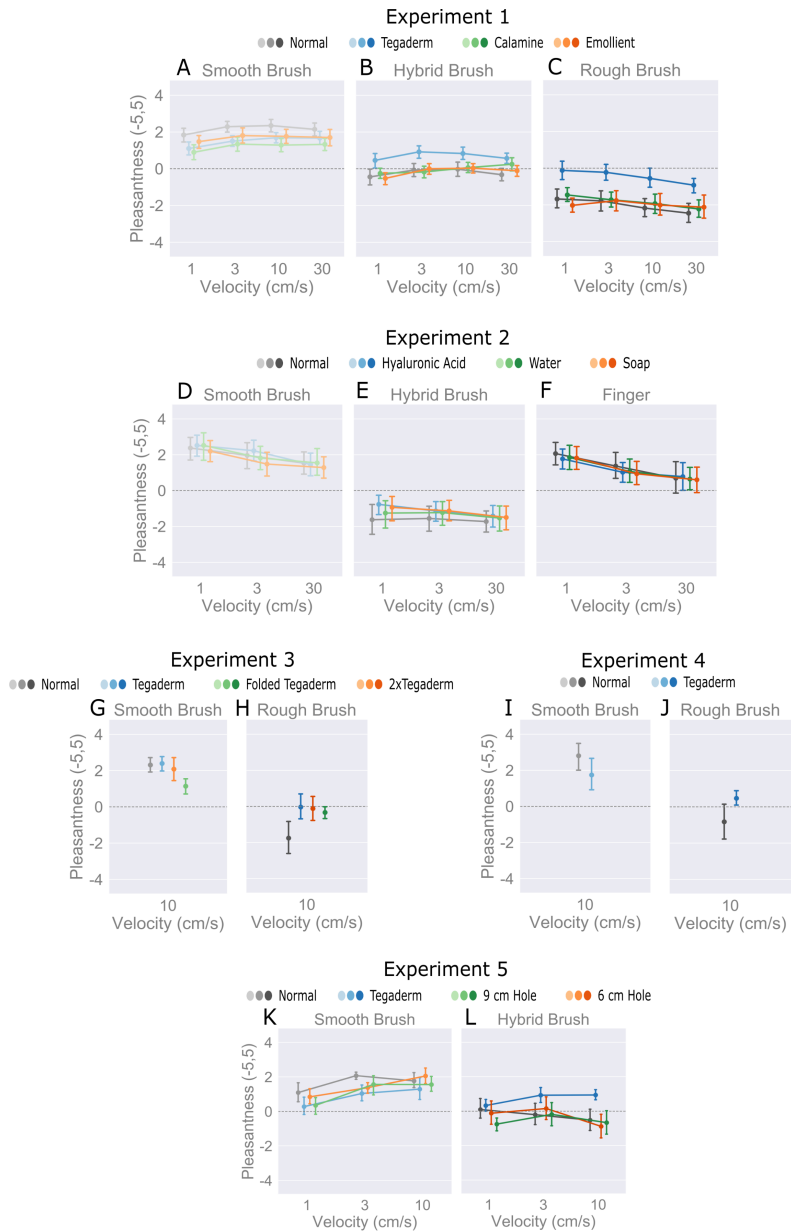

**Figure 2: Results of psychophysical experiments 1-5. (A-C)** Experiment 1 shows the relationships between brushes, velocities (1, 3, 10, 30 cm/s), and skin treatments meant to block direct contact, stiffen, and smoothen skin, respectively. **(D-F)** Experiment 2 investigated changes in frictional properties of the skin on pleasantness; with hyaluronic acid, washing with room temperature water (patted dry), and soap (patted dry) used to drastically increase friction, moderately increase friction, and decrease friction compared to the 'normal' condition at velocities of 1, 3, 30 cm/s. **(G-J)** Experiments 3 and 4 consider Tegaderm as a barrier and its adhesion to the skin when folded with no adhesion and with adhesion but two layers. **(K-L)** Experiment 5 shows the relationships between smooth and hybrid brush stimuli, accompanied by modulation of the skin's lateral movement, achieved by cutting holes of various sizes in the Tegaderm. In summary, brush stiffness and Tegaderm film modulated pleasantness, whereas other skin treatments, notably involving increases and decreases in friction, yielded little to no effect.

### F. Quantitative Measurement of Force during Brushing Procedures.

Perceptual differences were observed between the brush stimuli, though their forces and velocities, angles of contact, and location and area on the forearm, were controlled by a trained experimenter. To evaluate the force characteristics produced by each brush, we devised a test rig to measure normal force during brush strokes over a silicone-elastomer substrate (10 cm diameter, 60 kPa modulus, BJB Enterprises, Tustin, CA; TC-5005 A/B/C) lightly covered with baby powder to mimic the elastic and frictional properties of skin, Fig. 3A. Normal force data was captured via a uniaxial load cell (5 kg, 80 Hz, HTC Sensor TAL220, Colorado USA).

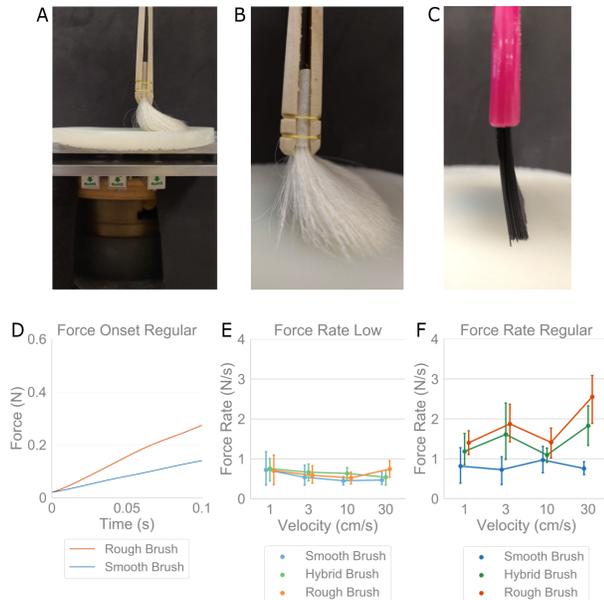

**Figure 3: Evaluation of force rate with brush stimuli**. (A) Experimental setup to collect force data from brush stimuli using a uniaxial load cell underneath a skin-like silicone-elastomer substrate, at two force levels with 'Low Force' meaning barely making contact and 'Regular Force' used in the psychophysical experiments, at velocities of 1, 3, 10, 30 cm/s. (B,C) Smooth and rough brushes in contact with the surface, respectively. (D) Force data over first 100 msec of contact onset at 'Regular Force' for an example trial per brush. The rough brush exhibits a higher force rate than the smooth brush. (E,F) Force rate at onset of contact, for all three brushes, again at two force levels and four velocities. Force rates at 'Low Force' are stable around 0.5 N/s for all stimuli, but at 'Regular Force' the force rate magnitude and variance increase significantly for the stiffer brushes.

Brush strokes were executed at velocities of 1, 3, 10, and 30 cm/s, and at two different force levels. In Fig. 3A-C, the experimental setup is shown with the smooth and rough brushes in contact with the silicone substrate, respectively. 'Regular Force' was the force (0.4 N) used in all prior psychophysical experiments, whereas 'Low Force' denotes a minimal level of contact between the stimulus and substrate, executed for comparative purposes. Brushing procedures were identical to Experiments 1-5 with each trial consisting of three separate, forward, back, and forward motions.

*Results.* Force rate over the first 100 msec of contact was analyzed due to its role as an efficient means in encoding object compliance, as opposed to other cues tied to stimulus velocity [25], [26]. This method was more appropriate due to the placement of the uniaxial load cell, as continuous force readings would not account for unavoidable torques produced during brush strokes. The rough brush has a faster increase in force than the smooth brush, Fig. 3D. Fig. 3E-F show the force rates across all brushes and velocities, highlighting their relationship with respect to using 'low' and 'regular' contact forces. In Fig. 3E, at the 'Low Force' level, peak force rates were consistent between brushes as well as the velocities. Likewise, for the smooth brush at 'Regular Force,' force rate remains relatively unchanged between velocities, Fig. 3F, as well as compared to its 'Low Force' level in Fig. 3E. However, for the hybrid and rough brushes, force rates at 'Regular Force' increase significantly over their 'Low Force' levels, as well as compared to the smooth brush at the 'Regular Force' level, Fig. 3F. They also exhibit larger trial to trial variability.

## IV. Discussion

This effort performs a series of psychophysical experiments to study the role of brush stiffness and skin treatments in encoding pleasantness at skin contact. While the relationship between brush velocity and pleasantness has been widely replicated, we do not yet understand how skin movements – e.g., lateral stretch, stick-slip, normal indentation – drive us to form such judgments. We take a distinct approach by 1) varying the properties of stimuli, by using brushes of distinct bristle stiffness and the human finger, and 2) utilizing skin treatments that isolate the underlying attributes of adhesion, friction, film thickness, and lateral mobility at the contact interface. Overall, the results indicate that a brush's stiffness influenced pleasantness more than any skin treatment. Velocity has been shown to have selective effects on pleasantness in earlier work, but more recent research suggests that the negative quadratic relationship between velocity and pleasantness ratings does not exist at the individual level [9]. Surprisingly, varying the skin's friction did not affect pleasantness. However, the application of thin film modulated pleasantness. Such barriers, though elastic and only 40 microns thick, inhibit the skin's tangential movement and disperse normal force.

First, we find that greater brush stiffness decreases pleasantness. Indeed, most prior works on pleasantness tend to use only a smooth brush and vary velocity, but changing brush stiffness decreases pleasantness much more, comparatively, than change in velocity. Work is still required to understand exactly why. A likely possibility, is a higher activation of c-nociceptors [27] in conjunction with c-tactile afferents when increasing brush stiffness. In alignment, in our instrumented force measurement experiment, Fig. 3, we find that differences between the brushes in their produced force rate at the onset of contact. Indeed, higher force rates may be less pleasant and their modulation may inform the dimension of valence. In Fig. 3, testing the stimuli at a low force level revealed a cross-velocity similarity for force rates, and for hand held stimuli [10]. Interestingly, the smooth brush's force rate did not vary with increased force application. However, such an increase was observed for the hybrid and rough brushes. Furthermore, since a low force rate shows high correlation with brush stiffness, and the smooth brush was the most pleasant of the stimuli, we can speculate that if force rate is controlled at a sufficient precision, a conventionally stiff stimulus might be made to be perceived as pleasant. That said, since these brushes are composed of different materials, factors other than just bristle stiffness are changing simultaneously, such as contact area and force concentrations on the skin. These factors need to be decomposed individually.

Second, skin treatments such as Tegaderm, attenuated the pleasantness of brush stimuli, while the modulation of friction played a minimal role. While initially it might seem intuitive to draw the conclusion that this is solely due to the presence of a direct barrier between skin and stimulus, there are likely more complex phenomena at play. Pleasantness perception has been strongly correlated across the range of velocities from 0.1 to 30 cm/s to the firing frequency of C-tactile afferents, with a lack of correlation to the firing patterns of Aβ afferents [3]. C-tactile afferents respond optimally to lateral brush strokes of 1-10 cm/s, but no systematic work has been done on the force ranges that either saturate the afferents or fail to evoke a response. Moreover, a comparison between 0.2 and 0.4 N indentation force on the responsiveness of C-tactile afferents to brushing revealed no consistent effect [3]. In addition to vertical inhibition of skin movement and modulation of force, Tegaderm film may also be effective in inhibiting lateral movement of the skin, though our attempt to simply cutting holes in the Tegaderm film did not attenuate pleasantness; therefore, its role as a direct barrier seems to be still required. It is important to note Tegaderm's side effect of immobilizing hair follicles during brushing. However, a prior study showed that depilation does not affect perception [28].

Finally, the finger as a stimulus was perceived to be close to the smooth brush in pleasantness, Fig. 2D and 2F. We do not know what exactly causes this similarity since the smooth brush and finger are quite different from each mechanically in both static and dynamic conditions. Perhaps there are ties to recent work finding that softness, as a psychophysical percept, comprises of five separate dimensions of granularity, deformability, viscoelasticity, furriness, and roughness in active, discriminative touch [29].

ACKNOWLEDGMENTS

This work was supported in part by grants from the National Science Foundation (IIS-1908115) and National Institutes of Health (NINDS R01NS105241). The content is solely the responsibility of the authors and does not necessarily represent the official views of the NSF or NIH.